# Creating an Optimal Portfolio of Crops Using Price Forecasting to Increase ROI for Indian Farmers

Authors:

Akshai Gaddam (akshai_g@yahoo.com),
Sravan Malla (sravanmalla1992@gmail.com),
Sandhya Dasari (sanjunriit@gmail.com),

Co-Authors:

Prof. D. Narayana
(darapaneni@gmail.com) - Program Director, Big Data & Machine Learning

Mukesh Kumar Shukla
(mukesh.shukla@gmail.com) - Faculty at Great Lakes, Big Data

*Great Learning*

**Abstract**

The Indian agricultural sector being in a constant phase of upgradation, has been on the road to modernization for the last couple of years. The fundamental source of livelihood for over 70% of the population living in rural parts of the country is still agriculture. The average Indian farmer, although has access to raw and trend data pertaining to crop prices, yield and demand (year-on-year) from Indian government and private websites, still struggles to make the right choices. They are constantly faced with the dilemma of choosing the right crop to address market demand and fetch them a decent profit. Since the process of shortlisting crops amongst the many suitable ones isn't completely scientific and usually dictated by area tradition, this project has aimed to address that issue by forecasting the price of those crops and uses that to create an optimal portfolio that the farmers can obtain to arrive at a data-driven decision for crop selection with optimal estimated ROI.

Keywords: Forecasting, Time-series, Machine learning, Big Data, Data Analytics, Agriculture, Algorithms

**1. Introduction**

The small to medium scale Indian farmers have very low risk appetite and limited capital for investment and get badly hit when they make uninformed decisions and end up not getting the optimal ROI. This is owing to multiple factors around agricultural infrastructure in the nation which is vastly different than that of developed countries. So, whenever a farmer ends up making wrong choices from choosing the crop to match the market demand and ends up in losses, the low risk appetite can drive them towards acquiring loans, turning them into debtors. Some farmers who are unable to pay these loans in time, even commit suicides (In 2014, the National Crime Records Bureau of India reported 5,650 farmer suicides). Despite there being multiple calls for modernizing the Indian Agricultural sector, the progress has been limited. The problem being multifold, not only is the agricultural infrastructure to be improved, but the poverty and ignorance of the farmer also need to be dealt with. Medium scale farmers who have access to smart phones, apps and internet have access to limited amount of historical price related data online and, the usage of that to arrive at a data-driven decision to invest in the right crop (in the right ratio) is not to full affect as they are not equipped or trained to deal with the raw data. Farmers mostly gather this information from cooperative societies and local markets manually and take a decision which isn't a very scientific or data driven approach. There currently exists no capability for using available raw data to provide forecast data to plan a portfolio of suggested crops for the farmer to plant in the coming years with an ROI estimation.

The following 2 problems are handled in this article:
1. **Forecasting crop prices for upcoming year**
2. **Creating an optimal portfolio of crops for farmers to use as guidance.**

**2. Problem Definition**

Farmers in India can be categorized into 3 types:

1. The **small / marginal-scale / landless farmer** who makes up for 74% of the farming community
    a. Typically has ownership of < 5 acres (marginal has less than 2.5 acres) of land
    b. Very low access to information / data
    c. Little to no risk appetite and resilience to loss – needs to borrow for buying inputs
    d. Dependent on rains, is low skilled, has poor quality soil
    e. Decisions are influenced by area tradition, his dietary needs, culture and geography
2. The **medium-scale** who makes up for 19% of the farming community
    a. Typically has ownership of < 25 acres of land that is partially irrigated
    b. Has medium risk appetite and resilience for loss - partially mitigated through contract farming
    c. Uses part of the land to grow his food and vegetables for his family
    d. Has access to partial information local suppliers, agriculture agents, radio, television and internet
    e. Decisions are influenced by area tradition, profitability and family consumption
3. The **large-scale farmer** who makes up for 7% of the farming community
    a. Typically has ownership of > 25 acres of land – well irrigated
    b. Has high risk appetite and resilience for loss
    c. Has great access to information, markets and capital
    d. Takes into consideration factors like markets, profitability, soil, crop rotation etc.

The project aims to address the optimal portfolio requirement of the small to medium scale farmers who have low-risk appetite and limited amount (or unable to use trend data) of data for decision-making. The optimal portfolio generated can solve for many problems as listed below:

1. Farmer can plan and use his limited resources more efficiently
2. Blends his practical knowledge with data / scientific methods
3. Incorporates factors that have impacted cost, yield and demand historically
4. Helps address and meet the market demand
5. Fetches the farmer the right kind of ROI for his efforts
6. Reduces risk for those with low risk appetite
7. Educates farmer on available choices backed by data to also address ignorance

## 3. Data Preparation

Agricultural data pertaining to retail commodity prices, previous year's demand, fetched price is available in certain websites such as:

- rpms.dacnet.nic.in
- www.agmarket.gov.in
- www.data.gov.in

The quality of data on each website varies. Although raw and trend data is available for most recent years, there are cases of incomplete or missing values in some. Users would have to manually collate information regarding multiple variable that correspond to the respective demography, time and year before attempting to analyze it. Creating a price forecasting model to then take into consideration all the input information required from the farmer like:
1. number of acres available for farming
2. storage capacity available
3. Investment capital at hand
4. Wishlist of crops pertaining to the region

To start with, the steps taken were:

1. Data on retail prices was collected for four main crops (Jowar, Rice, Maize and Urad) from https://rpms.dacnet.nic.in and proceeded for analysis.
2. Data was then consolidated for all the regions, calculating max, min, mode, median and average price and framed in a dataset for each crop/commodity, considering the Mean/Average for further analysis from multiple sites.
3. The final data set used for analysis and model building contained average retail prices across all the regions for last few year
4. Starting with linear regression for predicting the prices we tried implementing other forecast models such as time-series to forecast the retail prices in future months
5. Data analysis, Visualization, Model building and Validation are all done in Python whereas optimization portfolio logic is built in R

The extracted raw data contained data for 7 years (2001 -2018), in which it was observed that between 2001-2005, there were a lot of missing values and data was not recorded at many levels. Decision was then taken to consider data from 2006-2018 for further trend analysis. Although as expected, year-on-year prices indicate an overall increasing trend, sudden drops/hikes in prices were also noted. This was primarily owing to the impact of good/bad monsoons and other external/climatic conditions. For this project however, we decided to exclude external factors into consideration limiting the scope of current project to forecast based on past trends only. Enhancements to the data and model to incorporate external factors are in pipeline.

## 4. Standard algorithms used for forecasting crop prices

Since prediction of commodity prices was the first part of the problem we aimed to solve for (which is a continuous variable), we started with the linear regression model, followed by the naïve forecast method and then the more sophisticated time series models. Below is the complete list of Algorithms/ML Models used.
- Linear Regression
- Naïve Forecast Approach
- Simple Average
- Moving Average (MA)
- Simple Exponential Smoothing
- Holt's Linear Trend Method
- Holt's Winter Additive Method
- Holt's Winter Additive & Multiplicative Method
- ARIMA/X
- SARIMA/X

**Comparison of Algorithms – Table 1**

| Algorithm | Advantages | Disadvantages |
|---|---|---|
| Linear Regression | Lower complexity, easy to understand | Doesn't suit 'time dependent' data |
| Naïve Forecast Approach | Lower complexity, easy to understand | Predicts just based on last observed point, no seasonality/trend in consideration |
| Simple Average | Lower complexity, easy to understand | Predicts based on average value which doesn't suit for non-stationary series with seasonality & trend |
| Moving Average (MA) | Takes rolling mean (window size) into consideration | Predicts based on average value of sliding window which might not always work |
| Simple Exponential Smoothing | Weighted Averages are taken into consideration | Doesn't take seasonality/trend into consideration |
| Holt's Linear Trend Method | Best suits for data with trend and high variations | Doesn't suit for data with seasonality |
| Holt's Winter Method | Best suited for data with trend, seasonality and high variations | Complex, need to understand additive and multiplicative techniques as per the series |
| ARIMA/X | Best suited for data with trend and high variations and aims to describe correlations in the data | Seasonality is taken into consideration with some extra order parameters |
| SARIMA/X | Best suited for data with trend, seasonality and high variations | Complex in handling, tuninge and find best order parameters with low AIC/BIC values |

**Table 1** above lists the advantages and disadvantages of each one of the considered models.

## 5. Data Preparation

Weekly retail prices data was downloaded from the websites as csv file and are loaded as data frame on Jupyter after following the various steps of data cleansing.

The raw extract of data containing retail price for every region and date along with other attributes like "Per Unit", "Variety" etc. from which Maximum Price, Minimum Price, Modal Price, Median Price and Average Price (Mean) for every week across all regions was calculated and framed into a dataset for each of the four crops taken into consideration (Rice, Maize, Urad & Jowar).

The final data frame contained the following attributes, namely:

- Week Date (DD-MM-YYYY)
- Maximum Price
- Minimum Price
- Modal Price
- Median Price
- Average Retail Prices (weekly)

After going through the data in detail, it was decided to consider attributes like "Week Date" and "Average Retail Prices" for all future analysis.

- Total number of crops used for analysis is 4 (Jowar, Rice, Maize and Urad)
- Total number of years considered for analysis is 5 (2012 to 2016)

## 6. Exploratory Data Analysis and findings

- The date column in final data set used for analysis is parsed and created date column as index for time series analysis.
- Dropped all other attributes viz. High/Low/Median/Modal price retaining only "Mean Average Price".
- Checked for existence of stationarity using different stationarity test

Some key findings are
1. Constant increasing trend observed in the prices of Rice and Jowar year-on-year
2. Though there is an overall increase, the price of Maize seems steady post mid 2013
3. There is a substantial dip in prices of Urad post 2016 and is indicative that the trend is not stable
4. Existence of Seasonality is observed in all the 4 crop prices.

**7. Step by Step walk through of the solution**

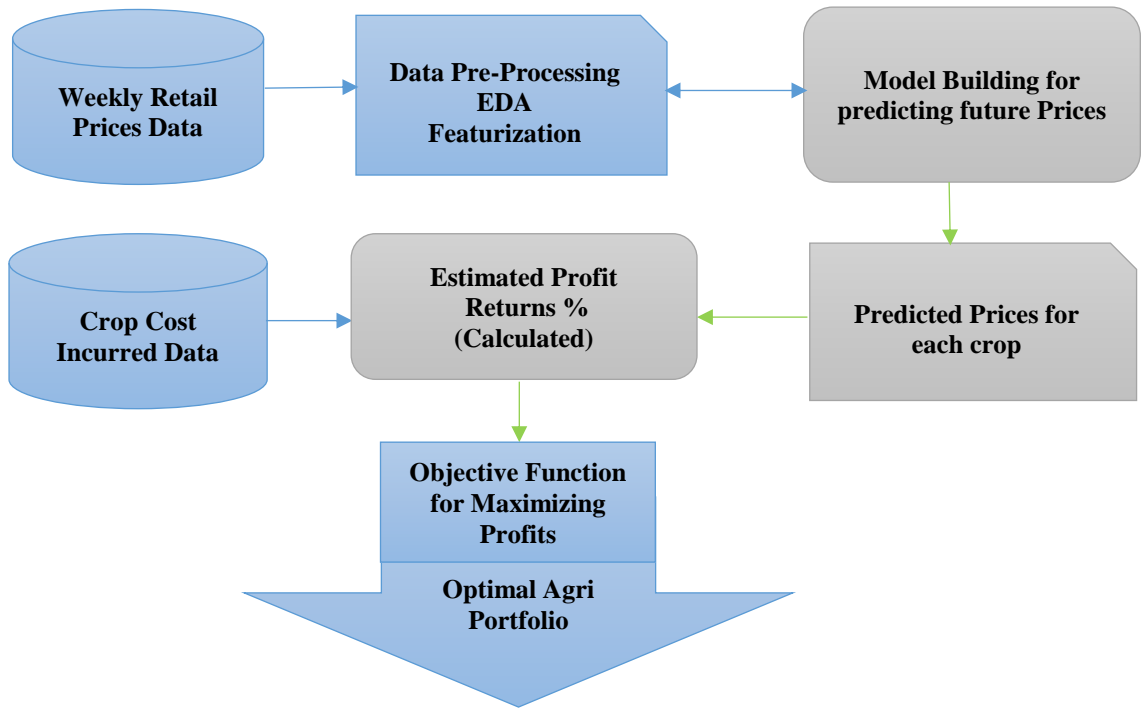

**Fig.1 - High Level Design Flow**

The above figure (Fig 1) is for representation purpose only, actual % varies based on data. The flow above indicates how after preprocessing of weekly data is done and loaded into the model for price prediction, the output of that is used to calculate estimated profit and in-turn, the final suggested optimal portfolio.

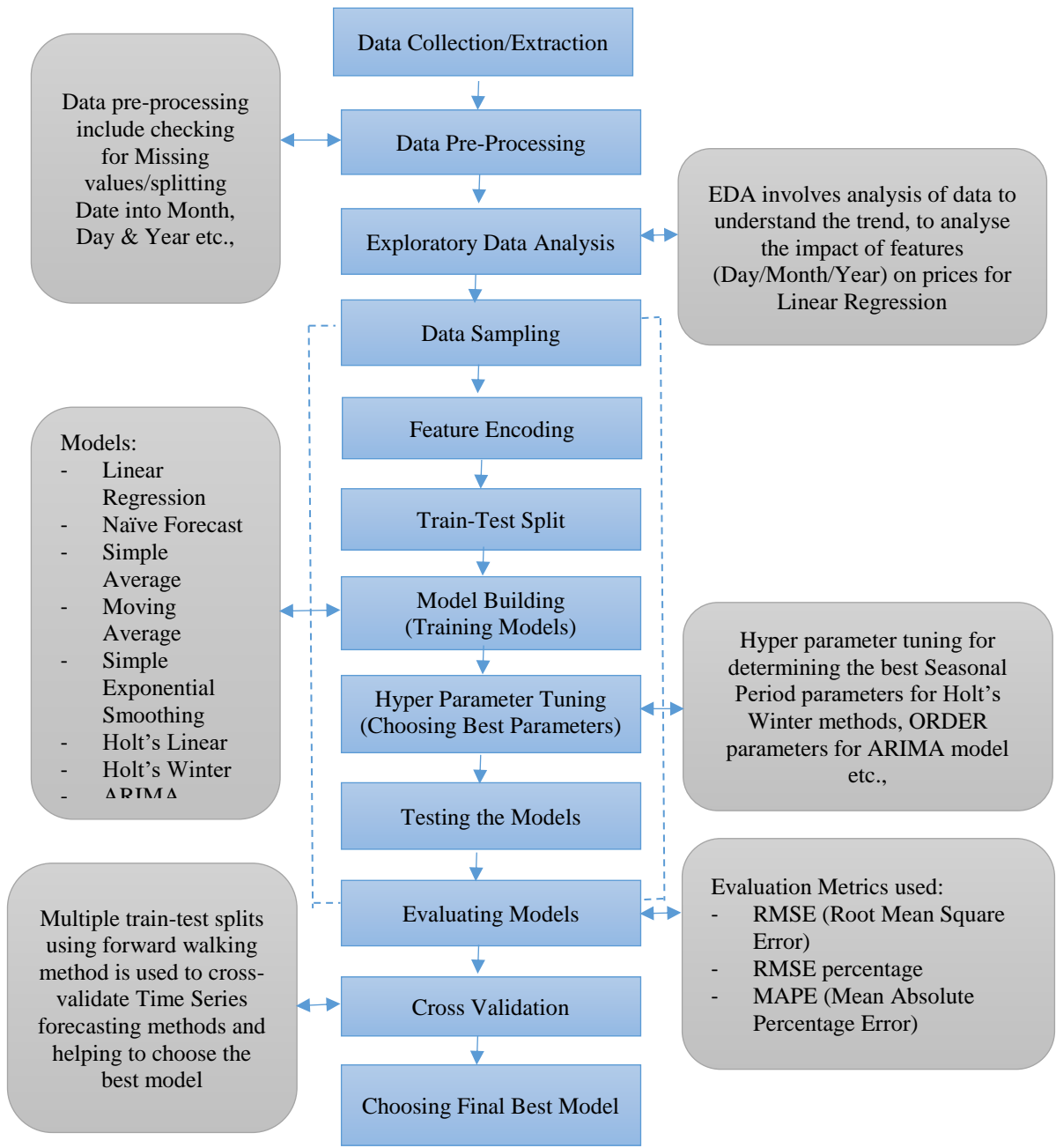

**Fig 2. Overall methodology workflow**

The above flow (Fig 2) is indicative of the final process considering all the data processing, EDA, Model building, Hyper parameter tuning, Evaluation metrics and data splits for time series analysis done.

## 8. Model evaluation and findings

### 8.1. Shortlisting the model

A comparison study was conducted on performance errors of all 9 shortlisted models using corresponding **RMSE (Root Mean Square Error), RMSEP (Root Mean Square Error of Prediction)** and **MAPE (Mean absolute percentage error)** values.

**JOWAR**

| | Models | RMSE | RMSEP | MAPE |
|---|---|---|---|---|
| 0 | Linear_Regression | 2.243456 | 8.168500 | 7.698133 |
| 1 | Naive_Approach | 0.737146 | 2.683976 | 2.179196 |
| 2 | Simple_Avg | 3.639938 | 13.253141 | 12.920236 |
| 3 | Moving_Avg | 0.806146 | 2.935206 | 2.335597 |
| 4 | Simple_Exp_Smth | 0.746390 | 2.717632 | 2.204313 |
| 5 | Hlt_Inr_Trnd | 1.111776 | 4.048015 | 3.467972 |
| 6 | Hlt_Wntr_Add | 1.508019 | 5.490750 | 4.846700 |
| 7 | Hlt_Wntr_Mul | 0.758887 | 2.763133 | 2.225635 |
| 8 | ARIMA/X | 1.647623 | 5.999052 | 5.332945 |
| 9 | SARIMA/X | 1.451450 | 5.284780 | 4.631501 |

Table 2 (a)

**MAIZE**

| | Models | RMSE | RMSEP | MAPE |
|---|---|---|---|---|
| 0 | Linear_Regression | 4.282575 | 17.098942 | 16.999248 |
| 1 | Naive_Approach | 1.591602 | 6.354753 | 5.398341 |
| 2 | Simple_Avg | 1.957323 | 7.814961 | 6.963147 |
| 3 | Moving_Avg | 1.577343 | 6.297822 | 5.339141 |
| 4 | Simple_Exp_Smth | 1.475740 | 5.892155 | 4.908368 |
| 5 | Hlt_Inr_Trnd | 3.819142 | 15.248603 | 12.389920 |
| 6 | Hlt_Wntr_Add | 2.252350 | 8.992908 | 8.686786 |
| 7 | Hlt_Wntr_Mul | 2.295128 | 9.163705 | 8.896446 |
| 8 | ARIMA/X | 2.724927 | 10.879755 | 10.679164 |
| 9 | SARIMA/X | 2.709851 | 10.819560 | 10.593110 |

Table 2 (b)

**RICE**

| | Models | RMSE | RMSEP | MAPE |
|---|---|---|---|---|
| 0 | Linear_Regression | 4.310115 | 10.006444 | 9.736291 |
| 1 | Naive_Approach | 2.702399 | 6.273941 | 5.438156 |
| 2 | Simple_Avg | 4.915528 | 11.411981 | 10.778200 |
| 3 | Moving_Avg | 2.035070 | 4.724656 | 3.991171 |
| 4 | Simple_Exp_Smth | 2.383764 | 5.534191 | 4.698359 |
| 5 | Hlt_Inr_Trnd | 2.493471 | 5.788889 | 4.859664 |
| 6 | Hlt_Wntr_Add | 2.427574 | 5.635901 | 4.759589 |
| 7 | Hlt_Wntr_Mul | 9.171035 | 21.291647 | 18.535839 |
| 8 | ARIMA/X | 1.131640 | 2.627236 | 2.213396 |
| 9 | SARIMA/X | 1.168414 | 2.712611 | 2.306518 |

Table 2 (c)

**URAD**

| | Models | RMSE | RMSEP | MAPE |
|---|---|---|---|---|
| 0 | Linear_Regression | 32.748299 | 22.142839 | 18.782887 |
| 1 | Naive_Approach | 12.417564 | 8.396165 | 6.997187 |
| 2 | Simple_Avg | 68.357766 | 46.220265 | 45.061996 |
| 3 | Moving_Avg | 12.497931 | 8.450505 | 7.243756 |
| 4 | Simple_Exp_Smth | 12.451427 | 8.419062 | 7.192559 |
| 5 | Hlt_Inr_Trnd | 11.663734 | 7.886462 | 5.813124 |
| 6 | Hlt_Wntr_Add | 36.928934 | 24.969586 | 19.831312 |
| 7 | Hlt_Wntr_Mul | 36.781821 | 24.870115 | 19.809760 |
| 8 | ARIMA/X | 44.354543 | 29.990429 | 24.264798 |
| 9 | SARIMA/X | 32.392298 | 21.902129 | 16.982014 |

Table 2 (d)

**Tables 2 (a, b, c, d)** above indicate the evaluation done on all considered models to get the RMSEP, RMSE and MAPE scores

Upon evaluation, it was observed that of all the models considered, using the **SARIMAX** and **Holt's Winter method** has resulted in lower error rates and hence a cross validation between the 2 methods needed to be done in order to choose the final method.

*8.2. Cross validation of models*

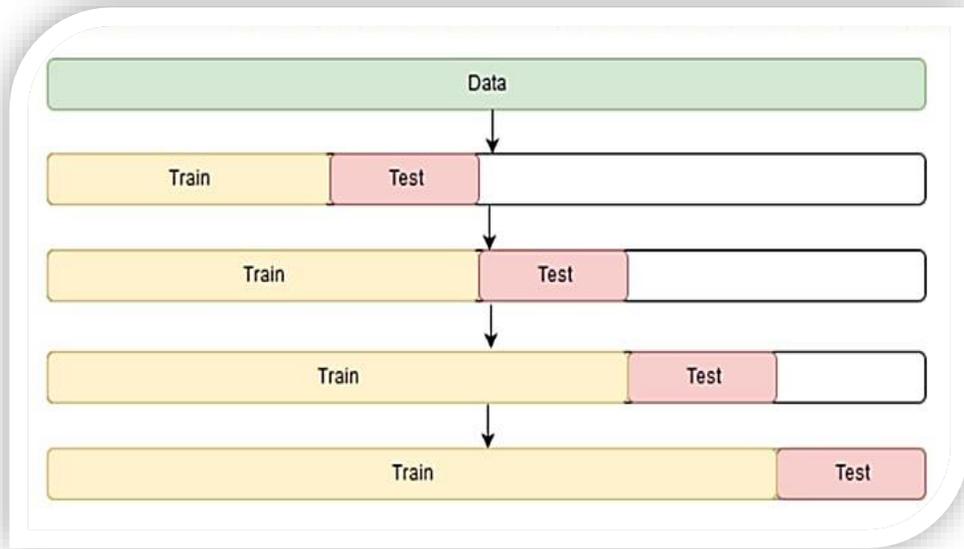

*Fig 3 – Cross validation of the Sarimax and Holt's winter models*

To choose the best of the 2 models, **Cross Validation on a Rolling Basis Method** was carried out on data of all 4 crops. Starting with a small subset of data for training (which we forecasted for the subsequent data points) and then checked the accuracy of the forecasted data points. The same forecasted data points are then included as part of the next training dataset and the process is repeated.

**JOWAR**

| | Models | RMSE | RMSEP | MAPE | | Models | RMSE | RMSEP | MAPE |
|---|---|---|---|---|---|---|---|---|---|
| 0 | SARIMAX_Split1 | 2.658493 | 11.653177 | 11.071971 | 0 | Holt_Winter_Split1 | 16.418092 | 71.966686 | 67.740477 |
| 1 | SARIMAX_Split2 | 1.552944 | 6.073714 | 5.185139 | 1 | Holt_Winter_Split2 | 1.003961 | 3.926588 | 2.680382 |
| 2 | SARIMAX_Split3 | 0.689140 | 2.561182 | 2.033007 | 2 | Holt_Winter_Split3 | 14.728071 | 54.736716 | 50.066323 |
| 3 | SARIMAX_Split4 | 1.548012 | 5.636368 | 4.969438 | 3 | Holt_Winter_split4 | 0.758887 | 2.763133 | 2.225635 |

**Table 4 (a)**

### RICE

| | Models | RMSE | RMSEP | MAPE | | Models | RMSE | RMSEP | MAPE |
|---|---|---|---|---|---|---|---|---|---|
| 0 | SARIMAX_Split1 | 3.203523 | 8.411212 | 7.495632 | 0 | Holt_Winter_Split1 | 2.433281 | 6.388855 | 4.445957 |
| 1 | SARIMAX_Split2 | 2.497223 | 5.869843 | 4.605054 | 1 | Holt_Winter_Split2 | 56.913192 | 133.777194 | 107.825847 |
| 2 | SARIMAX_Split3 | 3.733139 | 8.975924 | 8.289278 | 2 | Holt_Winter_Split3 | 53.482039 | 128.591722 | 107.750027 |
| 3 | SARIMAX_Split4 | 1.482994 | 3.442946 | 2.868817 | 3 | Holt_Winter_split4 | 10.289744 | 23.888862 | 20.770006 |

**Table 4 (b)**

### MAIZE

| | Models | RMSE | RMSEP | MAPE | | Models | RMSE | RMSEP | MAPE |
|---|---|---|---|---|---|---|---|---|---|
| 0 | SARIMAX_Split1 | 2.917494 | 12.530846 | 11.225773 | 0 | Holt_Winter_Split1 | 2.588676 | 11.118546 | 9.792324 |
| 1 | SARIMAX_Split2 | 1.469104 | 5.702787 | 4.571009 | 1 | Holt_Winter_Split2 | 1.592032 | 6.179973 | 5.012230 |
| 2 | SARIMAX_Split3 | 1.708709 | 6.649617 | 5.647061 | 2 | Holt_Winter_Split3 | 1.804494 | 7.022375 | 6.034302 |
| 3 | SARIMAX_Split4 | 2.709851 | 10.819560 | 10.593110 | 3 | Holt_Winter_split4 | 2.295128 | 9.163705 | 8.896446 |

**Table 4 (c)**

### URAD

| | Models | RMSE | RMSEP | MAPE | | Models | RMSE | RMSEP | MAPE |
|---|---|---|---|---|---|---|---|---|---|
| 0 | SARIMAX_Split1 | 2.240783 | 3.429078 | 2.122094 | 0 | Holt_Winter_Split1 | 1.682776 | 2.575158 | 1.700018 |
| 1 | SARIMAX_Split2 | 11.032883 | 14.081121 | 11.439679 | 1 | Holt_Winter_Split2 | 11.238903 | 14.344061 | 11.728586 |
| 2 | SARIMAX_Split3 | 33.458840 | 29.460036 | 20.222101 | 2 | Holt_Winter_Split3 | 44.443353 | 39.131745 | 28.340986 |
| 3 | SARIMAX_Split4 | 32.392298 | 21.902129 | 16.982014 | 3 | Holt_Winter_split4 | 36.781821 | 24.870115 | 19.809760 |

**Table 4 (d)**

**Tables 4 (a, b, c, d)** above indicate the comparison done between the Sarimax and Holt's Winter Method considering the RMSEP, RMSE and MAPE values

The cross validation resulted in **SARIMAX** proving to be the best fit for 3 of the 4 selected crops and so it was decided to go ahead with the **SARIMAX model** hence excluding the Holt's winter method.

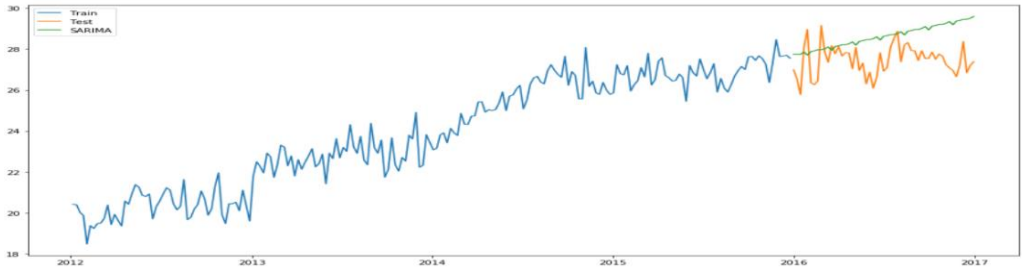

**Fig 4 (a). Jowar**

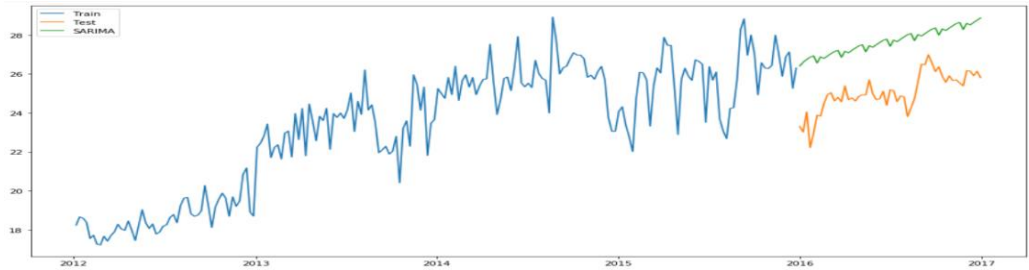

**Fig 4 (b). Maize**

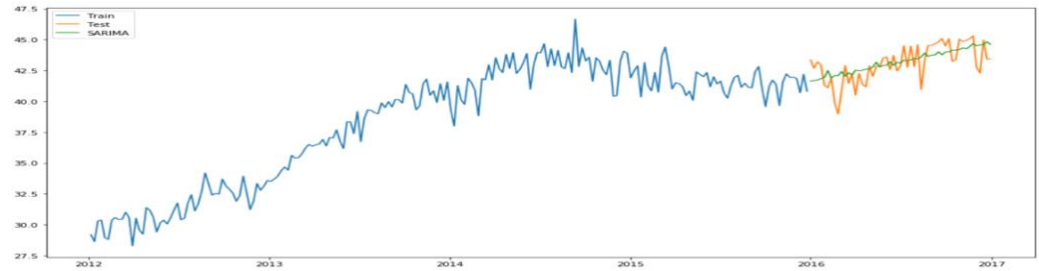

**Fig 4 (c). Rice**

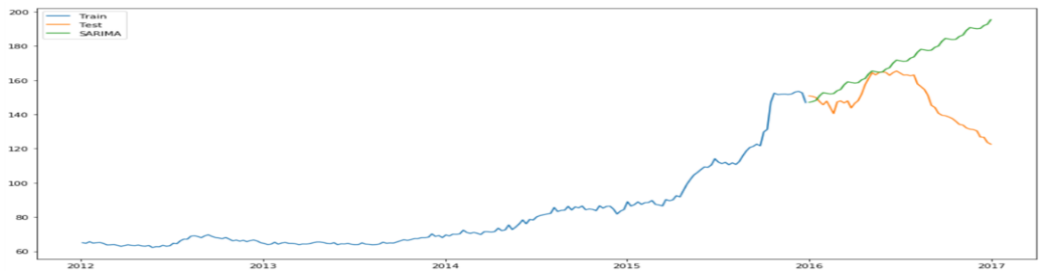

**Fig 4 (d). Urad**

**Figures 4 (a, b, c, d)** above indicate the SARIMAX plots done for all 4 crops

The plots above depict the comparison made between the predicted (green) curves for all 4 splits in cross validation (train-blue, test-orange). The results validated the decision to choose SARIMAX as the best model since it consistently performed for all types of variations in data and was the best fit for all the 4 crops.

## 9. Solving for the portfolio optimization problem

### 9.1. Sample Problem

Suppose a farmer has **20 acres** on which to plant four crops: RICE, MAIZE, JOWAR and URAD. To produce these crops, it costs the farmer (for seed, fertilizer, etc.):
- ₹ **22,500 /-** per acre for the RICE,
- ₹ **15,000 /-** per acre for the MAIZE,
- ₹ **7,000 /-** per acre for the JOWAR
- ₹ **10,600 /-** per acre for the URAD

The farmer has ₹ **2,00,000/-** rupees available for expenses. But after the harvest, the farmer must store the crops while awaiting favorable market conditions as per the forecasted prices. Let's say the farmer has storage space for **40,000 kgs**. Each acre yields an average of **3,000 kgs of RICE, 3000 kgs of MAIZE, 1500 kgs of JOWAR** and **350 kgs of URAD**.

If the net profit estimated from above forecasted prices and current cost price per kg (after all expenses have been subtracted) for **RICE is ₹ 4.50**/-, **MAIZE is ₹ 7/-**, **JOWAR is ₹10.34/-** and **URAD is ₹34.72/-**, how should the farmer plan the ratio of crops in his **20 acres** to maximize profit?

### 9.2. Sample Problem - Solution

Let's define the optimization function and the constraints.

$$\text{Max } G = (3000) * (4.5) W_1 + (3000) * (7.0) W_2 + (1500) * (10.34) W_3 + (350) * (34.72) W_4$$
$$= 13500*W_1 + 21000*W_2 + 15510*W_3 + 12152*W_4$$

Subject to below constraints formulated from above business case
$$22{,}500 W_1 + 15{,}000 W_2 + 7{,}000 W_3 + 10{,}600 W_4 \le 20{,}0000$$
$$3{,}000 W_1 + 3{,}000 W_2 + 1{,}500 W_3 + 350 W_4 \le 40{,}000$$
$$W_1 + W_2 + W_3 + W_4 \le 20$$
$$W_1 \ge 0$$
$$W_2 \ge 0$$
$$W_3 \ge 0$$
$$W_4 \ge 0$$

By using **LpsolveAPI**, we get the following results. To achieve the maximum profit of ₹ **3,47,383/-**, the farmer should plant:

**7.191 acres of Maize**
**12.123 acres of Jowar**
**0.684 acres of Urad**

    As and when forecasting is done for more crops, there would be more options available at the farmer's disposal to choose from crops that best suit his land, the region's climate and any other influencing factors.

## 10. Summary

Although other forecasting methods numerically reflected lesser error values their corresponding prediction lines were just linear without taking trend or seasonality into consideration. To choose the best final model between Holt's Winter (Add/Multiplicative) and SARIMAX, cross validation had to be performed to yield the best result.

## 11. Limitations

- Since external factors weren't taken into consideration, data was resampled to consider dates where high variations weren't observed. This made the model sensitive to external factors (i.e. heavy rainfall, floods, monsoons water supply, type of land etc.)
- Cost incurred by the farmer for selected crops (i.e. for seeds, fertilizer, land, ploughing etc.), was considered by taking inputs from a single farmer from a specific location. The actual cost can differ based on demographics of the farmer and the type of land he owns.
- The constraints formulated in the optimization problem logic is just a sample business case, the constraints would again be different based on farmer's location / demographics. So, this model should be further extended to be region specific for better accuracy.

## 12. Challenges Faced

- Like most data science / analytics project on open world problems gathering data of Cost incurred for cropping (for selected 4-5 crops).
- Lack of expertise in the agricultural field made it challenging to validate the gathered data.
- If other variables need to be considered such as temperature, rainfall, production, yield, area sown etc., data may or may not available for same location/region for which retail prices are currently available.
- Not all datapoints gathered so far were available for same specific location, average retail price across the locations was considered for analysis.

## 13. Enhancements planned (by Q1'2019)

Although SARIMAX was chosen as the final model, there is still scope for further fine tuning to improve the accuracy of model. We also would like to explore more on model evaluation part as there are quite different ways to cross-validate Time series models compared to normal Machine Learning models. Also, not limiting Time Series Analysis, application of Deep Learning models where LSTM should be best suited for this kind of data is planned as enhancement.

## 14. Conclusion

Big data analytics using ML / DL techniques can make a big difference in the Indian Agricultural scenario and in turn, in the life of the Indian farmer. Most of the work on this Agriculture analytical project demanded a lot of time to be spent on data gathering, checking for its relevance with the problem statement, pre-processing the data and re-sampling the data based on results. Applying the ML/TS/DL models took considerably lesser time than data gathering and EDA. Also, additional research and learning on Time-Series Analysis and Optimization techniques was needed as the project team was applying them for the first time. Also, the team would like to further extend the construct of the model by following through with the planned enhancements in the above section.

**Acknowledgements**

This work was towards partial fulfilment of Post Graduate Program Big Data and Machine Learning at Great Learning. We thank Great Learning for the opportunity and support.

**Acknowledgements**

This work was towards partial fulfilment of Post Graduate Program Big Data and Machine Learning at Great Learning. We thank Great Learning for the opportunity and support.